# Dynamic Software Updating in Java - Comparing Concepts and Resource Demands

Danijel Mlinaric

Department of Applied Computing
Faculty of Electrical Engineering and Computing
University of Zagreb
Croatia
danijel.mlinaric@fer.hr

Vedran Mornar

Department of Applied Computing
Faculty of Electrical Engineering and Computing
University of Zagreb
Croatia
vedran.mornar@fer.hr

**Abstract**

Dynamic software updating (DSU) is an extremely useful feature to be used during the software evolution. It can be used to reduce downtime costs, for security enhancements, profiling and testing the new functionalities. There are many researches and solutions on dynamic software updating regarding diverse problems introduced by the topic, but there is a lack of research which compare various approaches concerning supported changes and demands on resources. In this paper we are comparing currently available concepts for Java programming language that deal with dynamically applied changes and impact of those changes on computer resource demands.

*Categories and Subject Descriptors* D.3.2 [**Programming Languages**]: dynamic software updating

## 1. Introduction

State of the art software systems should support most real world processes. Business, industrial, healthcare and traffic are typical examples. Such processes frequently change over the time and those changes must be supported by software evolution. Software evolves to support changes in underlying processes (adaptive maintenance), because of maintenance in a term of bug fixes (corrective maintenance), to enhance performance and user experience (perfective maintenance), or to prevent problems in the future (preventive maintenance) [18]. Software evolution involves application of the new software version or patch to currently running software. Standard software updating procedure, in a form of stop/update/start steps, introduces downtime to the end users in the moment of change. Software downtime is not desirable since it introduces costs in business (e.g. banking/transaction software) and can negatively affect critical processes (e.g. software controlling power distribution).

To provide software updating without interruption or downtime, dynamic software updating is introduced. In addition to introducing the new functionality and improvements [21], dynamic software updating can be used to facilitate introduction of the new prototype functionality and to analyze software behavior [20]. Many researches on dynamic updating identified numerous problems not found in the standard updating procedure [7, 17, 19] and proposed different concepts to support dynamic updating [3, 7, 13, 17, 19, 30]. Supported software changes regarding complexity can be as simple as a change of a single or multiple instructions in low level programming languages [21], or more complex as changing the data structures (e.g. class inheritance) in higher level programming languages, like Java [11, 31].

Challenges of dynamic software updating such as how to preserve previous state of the program after the update [13, 19] and performing the update at a safe point in program execution [17, 27, 30], exist regardless of software environment and programming language and can negatively affect wider utilization. For every dynamic software updating concept there exists a requirement to support unanticipated software changes with minimal impact on performance and the end users' usage experience [8, 11, 13, 19]. Presently, there exist several concepts for Java programing language that use various techniques which differ in dynamic updating logic [1, 11, 20, 22, 31]. Some of the approaches can be found in development environments [9, 31], growing the acceptance of dynamic software updating. However, there is a lack of comparison for different approaches regarding the supported changes and demands on resources. In this paper we are comparing currently available concepts in Java programming language and measuring impact of dynamic software updating on computer resources.

The paper is organized as follows. Section 2. briefly describes currently available Java programming language concepts and introduces categorization. Section 3. talks about resource demands measurements methodology. Evaluation and results are described in section 4. Related work is presented in section 5. , and section 6.

## 2. Concepts

The placement of dynamic updating logic is crucial factor of every dynamic software updating approach. Dynamic updating logic can be defined as logic responsible for dynamic deployment. In Java programing language environment, dynamic updating can be executed on different environment hierarchy levels (Figure 1), therefore using different techniques.

Higher level programming language execution environment such as Java programming language relies on virtual machine, i.e. Java Virtual Machine – JVM, an intermediate level software execution environment that supports code portability and is responsible to execute and translate intermediate portable code (Java bytecode) to machine code. Concepts involving virtual machine modification are based on modifying JVM and performing the program change with assistance of modified JVM internal data structures, class metadata, stack and modified garbage collection mechanisms as seen in [28, 31].

On the other hand, programming environments such as JVM support specific API-s (*Application Programming Interface*) mechanisms e.g. JVMTI - *Java Virtual Machine Tool Interface* agent or Java agent. Those API-s can intercept access, modify program constructs (e.g. classes) and current program state. JVM agents also provide bytecode manipulation which all together can be utilized for dynamic software updating. Concepts utilizing JVM agent often involve intermediate objects between the previous and the new object version as "proxies" and wrappers in [11, 24]. Additionally, to cope with different class versions, different class loaders [11] or class renaming [24] can be used.

Aspect oriented programming (AOP) enables extending the program functionality on horizontal hierarchy level (*cross-cutting concern*) [16], e.g. for logging or security access. AOP uses weaving mechanism to load the new program code within an aspect during compile or load time, as with Aspectj [16]. Dynamic aspect oriented paradigm – DAOP enables weaving during the runtime, so it can be used for dynamic software updating. It is implemented as JVM modification [2, 5, 23] or JVM agent [1, 22].

Regarding the architecture, mechanisms and paradigms found in currently available approaches, we classify concepts in following three categories: modified intermediate level (modified JVM - mJVM), level between intermediate level and executed program (JVM agent - JVMa) and concepts based on DAOP (Table 1). Figure 1 show Java environment vertical hierarchy, where JVM agent and DAOP are located in the middle, although DAOP can be implemented both as VM modification and JVM agent.

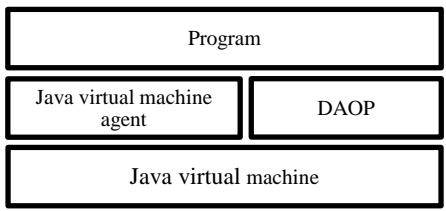

Figure 1 Java environment vertical hierarchy

Besides the aforementioned categorization, Java concepts can be compared by other key characteristics described in the following subsections.

### 2.1 Update timing

One of the crucial problem in dynamic software updating is to determine a proper moment in time to perform a dynamic update. It is extensively discussed in related literature [17, 27, 30]. However, Java concepts are mainly relying on internal JVM mechanisms. Dynamic update is executed on JVM *safe* points [12, 28, 31], with delayed or refused update when updating active code [12, 28]. At safe points, threads are stopped e.g. to perform garbage collection, which occur on loop endings, method calls, endings, etc. Besides the approaches relying on JVM defined update points, there are approaches with the programmer manually defined points such as [3, 24]. Further, in some cases a safe point may never be reached, e.g. in methods with long running loop. To prevent such specific cases, appropriate modification is required at the design time, e.g. using the functional decomposition of code inside the long running loop [19].

The moment of update completion is another point to consider. Modified JVM often use garbage collection mechanism to replace references to the modified objects, which implicitly performs updates atomically during the garbage collection [28, 31]. On the other hand, in lazy update approaches, objects are updated at the first access of a modified object [11, 12]. Atomic and lazy approach can be both found in modified JVM and JVM agent, while DAOP, due to join-point activation, performs aspects update atomically [20, 22].

### 2.2 Program adaptation

In various approaches to DSU, several types of program adaptation techniques can be found. Modified JVM and JVM agents introduce a mechanism to transfer object state from the previous to the new program version. There exist approaches with the programmer provided or automatically generated state transfer functions [12, 28], while other approaches provide mechanism to copy field values [9, 31]. DAOP concepts, except Dynamic ITD [2], generally do not provide such mechanisms. Moreover, in addition to state adaptation, manual program adaptation may be required to comply to conventions introduced by specific approach, e.g. programming language extensions presented in [3].

Current approaches perform program adjustments at compile, load or run time. In compile and load time, program adjustments are made with bytecode modification e.g. by inserting bytecode snippets, called "hooks", to enable runtime join-point activation in DAOP [23, 29], or to enable proxy or wrapper objects activation in JVM agents [11, 24]. Jooflux [22], for example, at load time replaces static method invoke instructions with the dynamic instruction which can be altered at runtime. DAOP only provides dynamic method body modification. To perform incremental dynamic updates and to extend changes supported by DAOP, extended DAOP approaches such as [6, 15] perform program adaptation in preparation steps, utilizing source analysis to convert program changes between versions into aspects. On the other hand, modified JVMs, do not require program adjustments in preparation steps, beside possible generated transformations functions as in [12, 28]. Moreover, compared with the JVM agents, they utilize smaller or negligible program adjustments, made during the runtime. DCEVM [31], a modified JVM approach, fills up objects which decrease in size with dummy fields to improve garbage collector execution performance. Jvolve [28] inserts *return barrier* instruction on stack to perform a delayed update, while Javelus generate code validity checks to enable lazy update. Furthermore, DAOP approaches Prose [20] and Hotwave [1], implemented as JVM agent use runtime bytecode inlining, while Steamloom [5] and Dynamic ITD [2] also utilize bytecode modification built as modified VM.

Therefore, program adjustments are categorized by the program life cycle where DSU is enabled: D - design, P - programming, C - compile, L - load and R - run time (Table 1).

Table 1 Java DSU classification

| Type | Change category | Program adjustments | Examples |
|---|---|---|---|
| mJVM | 3 | DPR | [12, 28, 31] |
| JVMa | 3 | DPLR | [11, 25] |
| DAOP (mJVM) | 2 | DPCLR | [2, 5, 23] |
| DAOP (JVMa) | 2 | DPCLR | [1, 22, 29] |
| Standard JVM | 1 | D | [15] |

### 2.3 Supported changes

Method body change is a common modification with the *HotSwap* functionality, which becomes part of Java Hotspot VM from version 1.4 [14]. Modified JVM, since it works on lower level of environment hierarchy, inside the JVM, can handle more complex changes [10], extending the existing *HotSwap* mechanism [31]. JVM agent approaches can support unanticipated changes, from adding/deleting class members to class hierarchy modification as seen in [11]. DAOP approaches are generally limited to class members changes because they operate at the method level, but there exist approaches that can provide hierarchy change by means of inter-type declarations, as seen in [2].

Therefore, changes supported by various approaches can be categorized into three basic levels, where each level supports all changes supported by lower levels:

1. Method body change;
2. Addition and deletion of methods, fields, constructors, classes;
3. Addition and deletion of interfaces including changes in class hierarchy i.e. change of the class/interface super type.

tion. To measure impact of dynamic updating logic on program execution in steady state, it is necessary to compare program execution time in unmodified environment to execution time in environment that enables dynamic updating, as found in [5, 12, 20, 28, 31, 31]. Those environments differ only in dynamic updating logic. Execution overhead is calculated as difference in percentage between execution time on the environment without and with dynamic updating.

### 3.2 Update duration

Dynamic update duration is one of the performance characteristics in dynamic software updating [7, 13]. It consists of preparation time and time to perform an update. Preparation time, for example, can include the time waiting for a safe point to occur and time to load class files into memory. Update duration may also include the time to restore program to normal execution. In more detail, DSU approach to perform a dynamic update may discard JVM optimizations, such as previously optimized compiled code. Consequently, program is executing slower because JVM requires some time to adapt to a current update as shown by authors of

```
Original:                                                          Modified:
public class Fact {                                                public class Fact extends Algorithm {  /* hierarchy change (AS) */
    private int counter = 0;       /* remove field (RF) */             private int result;                /* add field (AF) */
    private boolean count = false;
                                                                       public Fact(){                     /* add constructor (ACo) */
    public Fact(boolean count){    /* remove constructor (RCo) */          this.name = "fact";
        this.count = count;                                            }
    }                                                                  public int calculate(int n) {
                                                                           int res = 1;                   /* method body change (MB) */
    public int calculate(int n) {                                          if(n > 1) {
        if(count) counter++;       /* method body change (MB) */               for (int i = 1 ; i <= n; i++)
        if(n <= 1) {                                                               res *= i;
            return 1;                                                      }
        } else {                                                           return result = res;
            return calculate(n - 1) * n;                               }
        }                                                              public void getLastResult(){       /* add method (AM) */
    }                                                                      return result;
    public int getCallsCount(){    /* remove method (RM) */            }
        return counter;                                            }
    }                                                              public abstract class Algorithm{       /* add class (AC) */
}                                                                      protected String name;
Modification level: 1  2  3                                        }
```

Figure 2 *Multiple* test, original and modified class version with marked modification level

Figure 2 illustrates a Java program class modification example, where parts of code changed between two program versions are colored according to proposed categorization levels. Changes of types or names of fields or classes, as well as changes of signatures of methods or constructors, are absent from the proposed categorization. Those type of changes are actually made by multiple changes performed in sequence, e.g. deletion followed by addition, as in [28, 31]. Furthermore, real world changes frequently involve multiple basic changes at once, therefore they can be categorized as compound changes.

## 3. Measurements methodology

Software performance in DSU environments is inherently lower compared to environments without DSU functionality [26]. Time required to perform dynamic updates should be as minimal as possible as well as demands on resources before and after the update [7, 11, 13, 23]. To compare approaches, we focus on several resource demands measurements: steady state overhead, time duration necessary to perform dynamic update, modified state overhead and impact on memory usage.

### 3.1 Steady state overhead

Dynamic updating logic should not affect program execution when there are no dynamic updates i.e. during steady state execu-

DCEVM in [31]. Transient time duration is affected not only by dynamic update implementation but also by internal JVM mechanisms (e.g. JIT, OSR), therefore it is difficult to measure and it is out of the scope in this paper. Update duration is measured as time elapsed between the moment when dynamic update is invoked and the moment when the dynamic update logic resumes normal program execution.

### 3.3 Modified state overhead

Dynamic software updating can introduce performance overhead in method executions after the update [11, 23]. Depending on the approach, overhead can be persistent or temporary. Overhead can be persistent in approaches with inserted code snippets, e.g. join point activation in DAOP [1, 29] and proxy calls in JVM agents [11, 24]. Temporary execution performance degradation is a characteristic of some modified virtual machine approaches [12, 28, 31]. Optimizations are invalidated after the update and gradually recovered. As stated in previous subsection, time between invalidated optimization and recovery is denoted as transient time. To compare approaches by execution overhead, execution timing measurement is required on dynamically updated method and method updated with standard procedure. Two separate measurements are provided: short and long term. Short term execution refers to the first method call after the update. Long term execution involves multiple calls of the same method in a sequence.

### 3.4 Memory usage

Using DSU can increase risk of excessive memory usage, especially in approaches that allows different program version coexistence, such as [7, 28, 31]. To the best of our knowledge currently there is a lack of research which include memory usage measurements for Java approaches. Memory usage can be monitored before the dynamic update, during the dynamic update, immediately after the dynamic update and when modified code resumes execution. To determine dynamic updating impact on the memory usage, we measure difference in memory usage before and after the update.

## 4. Evaluation

Conforming to the proposed measurement methodology, we selected several java DSU approaches for the comparison [20, 22, 28, 31]. Main criteria for selection was public availability. Based on proposed categorization in section 1, Jvolve [28] and DCEVM [31] belong to modified JVM, while Prose [20] and Jooflux [22] belong to the DAOP category implemented as JVMa[1]. As far as we know, pure JVM agents are not currently publicly available. Javeleon [11] became commercial product [9] and Javadaptor [25] is not in public domain.

To measure resource demands according to the methodology described in previous sections, we are using macro and micro benchmark tests. In related literature [9, 20, 22, 31], macro benchmarks are used to evaluate steady state overhead in common software usage scenarios as they simulate real world applications (scientific computation, text processing, etc.). We used open source macro benchmark test suite DaCapo 2006MR2 [4][2], with tests that are executable on all selected approaches: *bloat*, *chart*, *hsqldb*, *jython*, *luindex* and *lusearch*.

Update duration, execution and memory overhead are measured with custom made micro benchmarks with developed test suite. Table 2 contains modification tests from the suite with associated categorization. Except method body and multiple modification, results for two tests, add/remove (A/R) are shown for each category/modification. Basic category tests perform simple modifications, e.g. in original version of Fact class from Figure 2, test *AddField (AF)* represents the addition of the field result . To perform changes on DAOP beyond method body modification, techniques from extended DAOP approaches are required, e.g. *client-supplier* from [6]. In Table 2 those tests are found under compound category because they are consisted of multiple changes, e.g. *AddFieldChangeMethod (AF+MB)* on Figure 2 consist of added field result and modified calculate method body that in modified version uses newly added field. Figure 2 presents *Multiple* test from Table 2 including various class member modifications (level 2) with supertype change (level 3). Table 2, under category *level* (Lev.) contains maximum level used for modification/test.

Evaluation tests are performed on Ubuntu 12.04.5 LTS kernel 3.13.0-32-generic virtual machine running on Hyper-V on the top of the Windows Server 2012 R2 Datacenter. Virtual machine was given 25% of the overall 2xIntel Xeon E5-2640 processor power, with up to 8 GB of system memory. We used the server compiler configuration on HotSpot VM for Jooflux, Prose and DCEVM and fast adaptive compiler configuration on Jikes RVM for Jvolve.

### 4.1 Benchmark tool

Although many approaches to dynamic updates rely on *HotSwap* mechanism interface [1, 20, 25, 29], a standard dynamic updating management interface across different approaches currently does not exist. Each selected approach has a specific implementation and interface. To perform micro and macro benchmark tests, we developed a benchmark tool (Figure 3). Furthermore, to perform measurements, we developed additional helper tools for each approach. Helper tools are used as interfaces to invoke dynamic update on specific approach. They contain micro benchmark tests which are supported by specific approach and contain shared logic to perform measures. Tests are manually adapted for each approach to achieve same program change across approaches. E.g. DAOP approaches requires program adjustments similar to those described in [6, 15].

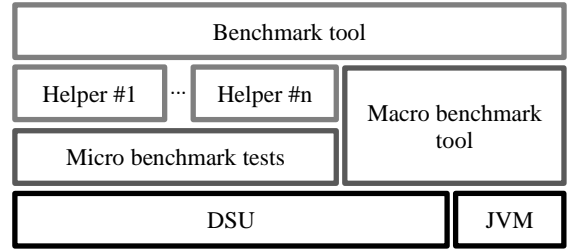

Figure 3 Measurement environment

In order to achieve "write once" tests for every subject approaches, universal format is required for description of changes. Furthermore, it is necessary to extended helpers with converters from the format describing changes to the format used by the specific approach. Developing such universal interface is out of the scope of this paper. Instead, we performed the tests after manual adaptation.

### 4.2 Results

For steady state overhead measurement, we run each test 50 times with DaCapo warmup option, in separate VM instances. Each test run was performed on two test cases, with and without dynamic updating. Overhead is calculated as a mean value of difference between times with and without DSU and expressed as a percentage. Results in Figure 4 present overhead across the test suite for selected approaches, except for *hsqldb* test for DAOP approaches and *jython* for DCEVM. In those cases, possible performance gain is negligible. For DAOP approaches, overhead in most tests is below the standard deviation, except for Prose in *lusearch* test. Moreover, results for *lusearch* demonstrate a high steady state overhead for all subject approaches except Jooflux. It shows that DSU approaches introduce significant overhead for multithreaded tasks processing large memory objects [4]. It can be noted that approaches which support more dynamic changes introduce more overhead in steady state. Regarding categorization in section 2., results show that mechanisms found in DAOP introduce lower overhead compared to modified JVM.

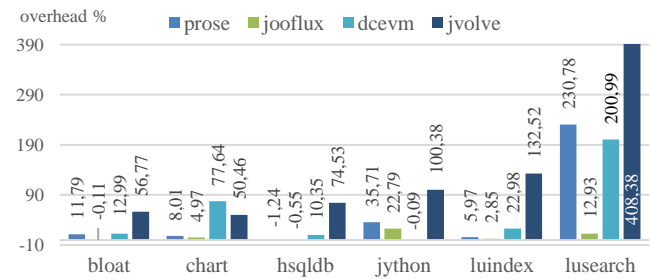

Figure 4 Steady state overhead

---

[1] Prose is implemented both as a modified Jikes RVM and Java agent. We used Java agent implementation.

[2] We have omitted the latest version, dacapo 9.12-bach, because tests from the test suite failed to execute on the Jvolve due to incompatibility problems related with Jikes RVM. Similar difficulties exist with specJVM2008.

As for dynamic update duration measurements, we performed 50 runs on separate VMs with 100000 test class instances. Results are presented as mean duration time in milliseconds to perform dynamic change for each test from Table 2. DAOP approaches Jooflux and Prose support level 2 changes only indirectly through compound changes ("**"). Jooflux lacks support for field change because it currently doesn't support the join-point functionality on the class field level. DCEVM supports all micro benchmark tests, while Jvolve partially supports *supertype* change, only as a compound change ("***"). Jvolve comparison tool lacks functionality to detect change of class *supertype* alone. Results show that values are lowest for the Jooflux, and highest for the Prose approach. Jooflux only performs call site switching using the *invokedynamic* [22], while Prose performs bytecode manipulation with method inlining technique [20]. DCEVM and Jvolve show similar results, as both are based on the modified JVM garbage collection mechanism. Execution times of different tests within the same approach do not vary significantly. Although, we can observe that DCEVM have higher values for the *AddField* and *AddSupertype* test because dynamic logic involves memory and class hierarchy rearrangement. On the other hand, Prose results show that field updates perform slower, which shows that join-point activation is slower for class fields.

Table 3 show differences in memory usage before and after the dynamic update. Setup of measurements is the same as for measuring durations and values denote mean difference in megabytes (MB). Only method body change and compound tests were measured since they are supported by the most of the approaches. It can be observed that some tests results have decrease of memory consumption. Jvolve, e.g. performs garbage collection after every update, followed by transformation function, if required. DCEVM activates modified garbage collector only in the case of adding of a field, to perform memory rearrangement. On the other hand, DAOP approaches introduce low memory usage, with exception of Prose for *AddField* test, which can be related to field join-point activation.

Table 2 Dynamic update duration with modification categorization

| Type | Lev. | Test/Modification | DSU duration [ms] | | | |
|---|---|---|---|---|---|---|
| | | | Prose | Jooflux | DCEVM | Jvolve |
| Basic | 1 | Method body (MB) | 685,85 | 3,07 | 63,49 | 52,59 |
| | 2 | Method (M*) | ** | ** | 59,11/65,23 | 117,56/114,02 |
| | | Constructor (Co*) | ** | ** | 81,64/63,72 | 116,54/110,53 |
| | | Field (F*) | ** | NA | 62,51/62,19 | 111,79/116,57 |
| | | Class (C*) | **** | **** | **** | **** |
| | 3 | Supertype (S*) | NA | NA | 84,75/78,99 | *** |
| | | Interface (I*) | NA | NA | 60,78/61,78 | 106,24/102,02 |
| Compound | 2 | M* + MB | 777,73/699,23 | 3,61/5,29 | 60,72/59,68 | 46,45/44,10 |
| | | F* + MB | 875,31/676,30 | NA | 75,41/65,28 | 46,45/44,098 |
| | | Co* + MB | 675,44/677,97 | 4,12/2,82 | 60,83/60,92 | 46,54/47,41 |
| | | C* + MB | 684,34/684,32 | 3,53/5,97 | 62,36/95,64 | 44,15/44,74 |
| | 3 | Multiple | NA | NA | 68,63 | 51,81/131,50 |

\* (A/R) Add/Remove test/modification  \*\*\* partially supported
\*\* indirectly supported  \*\*\*\* cannot be performed alone

Short term overhead, is measured over 50 separate VM runs which perform a single method call. Long term overhead is measured over 1000 multiple calls on same VM instance. Figure 5 contain results expressed as mean time in milliseconds in environment without DSU (standard) and environment with DSU. Here we performed *AddChangeMethod* (AM + MB) test, where non recursive method from Figure 2 is added to class and used from a method body which was changed in another class. Values show that short term overhead is lower for DAOP approaches. In long term execution, times are almost equal to times in environment without DSU. Jvolve is an exception, as it introduces a long term modified performance overhead, meaning that some JVMs optimizations must be discarded in order to perform dynamic updates.

## 5. Related Work

Our work is focused on comparing resource demands in different Java based approaches to DSU, categorized by concepts and supported changes. Authors in [26] introduce various metrics of the dynamic software updating features regardless of environment.

In related literature [9, 12, 20, 28, 31], steady state execution times are for the particular DSU approach, while we compare various DSU approaches. In [12, 28, 31], modified JVM approaches are evaluated by comparing modified garbage collection duration measured with micro benchmark tests performing class field changes. In contrast, we developed tests for various supported changes to evaluate and compare different approaches across those changes. DAOP approaches [1, 20, 22] evaluate dynamic aspect weaving and woven code performance, while our work compares the performance of the program before and after changes in program code.

In [8], authors introduce a quantitative cost-benefit model to estimate gain when using the dynamic updating compared to other updating techniques in different dynamic updating approaches. This comparison is expressed by a single revenue value calculated on the basis of the estimated model parameters. We do not attempt to make such an estimation because the parameters significantly differ across application domains.

Table 3 Difference in memory usage

| Test | Prose | Jooflux | DCEVM | Jvolve |
|---|---|---|---|---|
| Method Body | 0,13 | 0,41 | 1,18 | -5,66 |
| M + MB (A/R) | 0,06/0,13 | 0,97/1,00 | 1,28/1,30 | 2,81/2,82 |
| F + MB (A/R) | 4,64/1,02 | NA | -0,49/1,23 | 2,76/2,13 |
| C + MB (A/R) | 0/-0,1 | 0,48/-0,1 | 1,18/1,18 | -5,67/2,46 |
| Multiple | NA | NA | 1,25 | 1,87 |

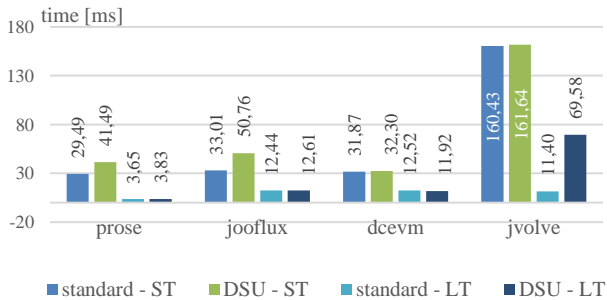

Figure 5 Short term (ST) and long term (LT) execution time

# 6. Conclusion

In extensively researched area of dynamic updating, couple of Java approaches [9, 31] matured enough to be used in development environments. To enable wider utilization and bring dynamic updating to production environments there is a need for standardization and dedicated benchmarks. We provided a brief overview of the existing Java approaches and categorized them on the basis of involved Java environment mechanisms. Various measurements and modification tests described here provide better understanding of internal workings that could lead to improvement of existing and development of new dynamic updating approaches. We created several test cases in the form of dynamic modification tests belonging to various categories of changes. We also created a benchmark tool to perform the measurements. Our main focus was to show how a particular modification affects the demands on computer resources. Evaluation results show that modified JVM (mJVM) approaches provide more complex dynamic changes (level 3) than DAOP approaches (level 2). This comes at a price of an overhead in steady state execution and memory usage. On the other hand, modified state overhead and change duration depend on particular approach mechanisms. Approaches using the internal JVM mechanisms, such as Jooflux, introduce low resource demands and do not require running environment modification such as mJVM, but support fewer types of changes. Provided that they expand the number of supported change types, those approaches could emerge as a simple and effective way to add dynamic updating to long running software systems.

Future work may involve introducing new categories for comparison, e.g. modification format or access flags modifications, and new measurements, e.g. steady state memory overhead or update duration in relation to number of object instances, as in [28, 31]. Furthermore, helper tool logic can be extended in flexibility by introducing universal change format to enable "write once" tests for all supported approaches.

## Acknowledgments

We would like to gratefully thank Ms. Marija Katic for her support, providing the materials during the paper preparation and advices during the research work. Without her help this research would not be possible.

## Appendix

Selected approaches source location and used JVM version:

- Prose: https://sourceforge.net/projects/jprose JDK 1.5.0_22-b03
- Jooflux: https://github.com/dynamid/jooflux JDK 1.7.0_85
- DCEVM: https://hg.java.net/hg/dcevm~repository JDK 1.7.0
- Jvolve: https://bitbucket.org/suriya RVM 3.0.1